\documentclass[showpacs,aps,amssymb,floatfix,prd,amsmath,preprintnumbers]{revtex4}
\usepackage{epstopdf}
\usepackage{color}
\usepackage{graphicx}  
\usepackage{dcolumn}   
\usepackage{bm}
\usepackage{hyperref}
\begin{document}

\title{\bf Unified Dark Fluid  and Cosmic Transit Models in Brans-Dicke Theory}

\author{Sunil K Tripathy\footnote{Department of Physics, Indira Gandhi Institute of Technology, Sarang, Dhenkanal, Odisha, INDIA, 769146, Email: tripathy\_sunil@rediffmail.com}, Sasmita Kumari Pradhan \footnote{Department of Physics, Centurion Institute of Technology and Management, Bolangir, Odisha, INDIA, Email: sasmita@gmail.com}, Zashmir Naik\footnote{School of Physics, Sambalpur University, Jyotivihar, Sambalpur, Odisha, INDIA, 769019, Email: zashmir@gmail.com}, Dipanjali Behera\footnote{Department of Physics, Indira Gandhi Institute of Technology,Sarang, Dhenkanal, Odisha, INDIA, 769146, email:dipadolly@rediffmail.com} and B. Mishra \footnote{Department of Mathematics, Birla Institute of Technology and Science-Pilani, Hyderabad Campus, Hyderabad-500078, India, E-mail:bivudutta@yahoo.com }}

\affiliation{ }

\begin{abstract}
Some dark energy cosmological models are constructed in the framework of a generalised Brans-Dicke theory which contains a self interacting potential and a dynamical coupling parameter. The dark sector of the universe is considered through a unified linear equation of state. The parameters of the unified dark fluid have been constrained from some physical basis. Since the universe is believed to have undergone a transition from an early deceleration to a late time acceleration, the deceleration parameter should have a signature flipping behaviour at the transition redshift. We have used a hybrid scale factor to simulate the dynamical behaviour of the deceleration parameter. Basing upon the observational constraints on the transition redshift, we have constructed four different transitioning dark energy models. The constructed models are confronted with observational data. For all the models, the behaviour of the dynamical scalar field, Brans-Dicke parameter, Self interacting potential are investigated. Also, on the basis of the generalised Brans-Dicke theory, we have estimated the time variation of the Newtonian gravitational constant.


\end{abstract}


\maketitle

\section{Introduction}\label{I}
A lot of observations during the last two decades have confirmed the late time cosmic acceleration \cite{S.per 1998, A.G 1998, R.Knop 2003, A.G 2004, A.G. 2007, D.N 2007, A.Blan 2007, D.J 2005, M.Sulli 2011, N.Suzuki 2012, C.R 2004, S.W 2004, S.P 2004, Cole2005}. These observations have developed a curiosity among the cosmologists to explain this late time dynamics. In the ambit of General Relativity  (GR), it becomes difficult to explain such a bizarre issue. Therefore, an unusual and exotic dark energy (DE) form with negative pressure has been conceptualized. The most intriguing thing is that, so far we do not exactly the nature and origin of this exotic energy source. The contribution of DE as compared to the baryonic matter and dark matter in providing an anti gravity effect to drive apart the universe for acceleration is the maximum. Recent Planck data estimates a lion share of $68.3\%$ in favour of DE \cite{P.A.R 5076,P.A.R 5082, P.A.R 5084}. The late time cosmic dynamics and the consequent dark energy is understood through a dark energy equation of state parameter $\omega_D=\frac{p}{\rho}$ where $p$ is the DE pressure and $\rho$ is the dark energy density. Experimental determination of this parameter has so far remained as a challenge to astronomers and cosmologists. Accurate knowledge of $\omega_D$ at the present epoch and its evolution history will definitely provide a good idea about the cosmic evolution leading to the late time cosmic speed up. In $\Lambda$CDM model, a cosmological constant represents the DE with an equation of state $\omega_D=-1$. In literature, $\omega_D$ is considered either as a constant or an evolving quantity. Time dependent DE equation of states can be obtained from canonical scalar field models such as quintessence field ($-\frac{2}{3}\leq \omega_D\leq -\frac{1}{3}$)\cite{Ratra1988, Sahni2000}, phantom fields  ($\omega_D < -1$) \cite{Caldwell2002}, tachyons \cite{Sen2002} and quintom fields \cite{Zuo2005, Urban2009}.  IN phantom fields, $\omega_D$ evolves in a phantom region with $\omega_D<-1$. In the quintom fields, $\omega_D$ evolves from a quintessence region to a phantom region through the phantom divide at $\omega_D=-1$. The DE equation of state parameter has been constrained from different observational data in recent times. The 9 year WMAP survey constrained  the DE equation of state parameter  from CMB measurements as $\omega_D=-1.073^{+0.090}_{-0.089}$ \cite{Hinshaw13}. A combination of the CMB data with Supernova data suggests $\omega_D=-1.084\pm 0.063$ \cite{Hinshaw13}. While Amanullah et al. have constrained it as  $\omega_D=-1.035^{+0.055}_{-0.059}$ \cite{Amanullah2010}, Kumar and Xu constrained it from a combined  analysis of the data sets of SNLS3, BAO, Planck, WMAP9 and WiggleZ constrained as $\omega_D=-1.06^{+0.11}_{-0.13}$ \cite{Kumar2014}. Moreover the recent Planck 2018 results constrained the DE equation of state parameter as $\omega_D=-1.03\pm 0.03$ \cite{Planck2018}. Amidst the theoretical explanation and justification for the existence of the exotic DE form, it has been widely accepted that the universe has undergone a transition from a decelerated phase at an early phase of cosmic evolution to an accelerated phase at late times. The transition might have occurred at a redshift $z_{da}\sim 1$ with the belief that the cosmic acceleration is a very recent phenomenon. Busca \cite{Busca13} constrained the transition redshift as $z_{da}= 0.82\pm 0.08$;  Farooq and Ratra constrained this as $z_{da}=0.74\pm 0.05$ \cite{Farooq13}, Capozziello et al. obtained a constraint on this parameter as $z_{da}= 0.7679^{+0.1831}_{-0.1829}$ \cite{Capo14}.  Reiss et al. derived kinematic limits on the transition redshift as $z_{da}= 0.426^{+0.27}_{-0.089}$ \cite{Reiss07}. While Lu et al. constrained this parameter to be $z_{da}=0.69^{+0.23}_{-0.12}$ \cite{Lu11}, Moresco et al. obtained the constraint $z_{da}=0.4\pm 0.1$ \cite{Moresco16}. Of late, there have been a growing interest in the dark degeneracy modelled through a unified dark fluid (UDF) where the contributions of both the dark energy and dark matter  are considered through a single DE equation of state. The single UDF equation of state may be linear or non linear in the energy density \cite{Anand06, Balbi07, Xu12, Liao12, SKT15}. The single UDF is able to explain most of the recent observational data.

Also, there have many attempts in recent times to modify the geometrical action by considering an arbitrary function of the Ricci Scalar and other variants of it including a bit of matter field coupled to the geometry \cite{Caroll2004, Nojiri2007, Harko2011, Nojiri2005, Linder2010, Myrza2011}. The motive behind such modification is to avoid the need of the exotic DE form about which we do not have a clear knowledge. These geometrically modified gravity theories provide ghost free models. It is worth to mention here that, additional dynamical degrees of freedom in the form of scalar fields are required to explain DE in the framework of GR. In some cases, these scalar fields are associated with unusual matter Lagrangian with negative kinetic energy which provide ghost models.  In the context of modification of GR, scalar tensor gravitation theories have played a major role. Particularly the Brans-Dicke (BD) theory is considered as a successful scalar-tensor theory to investigate the cosmic behaviour at different epochs such as the cosmic coincidence, inflation, and the cosmic acceleration \cite{Brans1961,Banerjee2001, Bert2000}. In the  BD theory, the gravity is mediated through a scalar field $\phi$ and the scalar field is coupled to the geometry through a coupling constant called Brans-Dicke parameter $\omega$. One interesting aspect of the BD theory is that it favours a time variation of the Newtonian gravitational constant and this time variation is associated with the time evolution of the BD scalar field. The BD theory has passed the experimental tests from solar system \cite{B.Bertotti 2003}. The theory is also tested against the CMB data  and large scale structure \cite{Wu2010, Wu2013}. Moreover, the BD theory has emerged as a low energy limit of many quantum gravity theories such as superstring theory or Kaluza-Klein theory \cite{SKT15}. Over a period of time, there has been a controversy in estimating the value of the BD parameter. While solar system experiments predicted a large value i.e. $\omega > 40,000$ \cite{B.Bertotti 2003}, it may be less than 40,000 on a cosmological scale \cite{V.Acqu 2007}. In a recent work, the BD parameter is constrained to be in the range $0.0014 < \frac{1}{\omega} < 0.0024$  or  $417 < \omega < 714$ \cite{H.Ala 2014}. Even negative values of the BD parameter such as $\omega < -120$ have been obtained from WMAP and SDSS data \cite{Wu2010}. Since the BD theory has a dynamical framework as compared to GR, it is interesting to investigate dark energy models in this framework.

In the present work, we have constructed some dark energy models with a signature flipping behaviour of the universe from early deceleration to late time acceleration in the frame work of generalised BD (GBD) theory.  In the GBD theory, there is a self interacting potential and a dynamical BD parameter that varies with the scalar field. This article is organised as follows. In Section II, the basic field equations and some dynamical properties of the universe are obtained for an anisotropic Kantowski-Sachs metric. The metric has different expansion rates along different spatial directions. In Section III, we have considered a UDF equation of state that unifies the dark sector of the universe. One part of the UDF describes the dark matter and the other part describes the dark energy. In fact, the UDF equation of state is a linear equation of state representing a barotropic fluid containing two adjustable parameters. The parameters of the UDF have been constrained using different observational results of some recent works.  In Section IV, we considered a hybrid scale factor (HSF) to simulate a signature flipping behaviour of the deceleration parameter. Four different HSF models have been constructed and compared with the observational $H(z)$ data. The evolutionary behaviour of the BD scalar field, BD parameter, Self interacting potential and the time variation of Newtonian gravitational constant have been investigated using the HSF models with UDF. At the end, the summary and conclusion of the work are presented in Section V.

\section{Field Equations in Generalised BD theory}
We consider here the generalised Brans-Dicke (GBD) theory with a self interacting potential.  The gravity is mediated through a dynamical scalar field and the BD parameter is considered as a function of the scalar field $\phi$. The action for the GBD theory in a Jordan frame is given by
\begin{equation}
S=\int d^{4}x\sqrt{-g}\left[\phi R - \frac{\omega(\phi)}{\phi}\phi^{,\mu}\phi_{,\mu}-V(\phi)+\mathcal{L}_{m}\right],\label{eq:1}
\end{equation}
where $\omega(\phi)$ is the modified BD parameter, $V(\phi)$ is the self-interacting potential, R is the scalar curvature and $\mathcal{L}_{m}$ is the matter Lagrangian. We chose the natural unit system: $8\pi G_{0}=c=1$, where $G_0$ is the Newtonian gravitational constant at the present epoch and $c$ is the speed of light in vacuum.  The field equations for the GBD theory are obtained as \cite{SKT15, SKT20}
\begin{eqnarray}
G_{\mu\nu} &=& \dfrac{\omega(\phi)}{\phi^{2}}\left[\phi_{\mu}\phi_{\nu}-\frac{1}{2}g_{\mu\nu}\phi_{,\alpha}\phi^{,\alpha}\right]+\frac{1}{2}[\phi_{,\mu;\nu}-g_{\mu\nu}\Box\phi],\label{eq:2}\\
\Box\phi &=& \dfrac{T}{2\omega(\phi)+3}-\dfrac{2V(\phi)-\phi\frac{\partial V(\phi)}{\partial\phi}}{2\omega(\phi)+3}-\dfrac{\frac{\partial\omega(\phi)}{\partial\phi}\phi_{,\mu}\phi^{,\mu}}{2\omega(\phi)+3}.\label{eq:3}
\end{eqnarray}  
In the above equations, $\Box$ is the d'Alembert operator, $T=g^{\mu\nu}T_{\mu\nu}$ is the trace of the energy momentum tensor 
\begin{equation}
T_{\mu\nu} = -\frac{2}{\sqrt{-g}}\frac{\delta\left(\sqrt{-g}\mathcal{L}_m\right)}{\delta g^{\mu\nu}}.\label{eq:4}
\end{equation}
The GBD theory reduces to the usual BD theory for a constant BD parameter $\omega$ and to GR in the limit of a constant scalar field and an infinitely large Brans-Dicke parameter $\omega$.

For a Kantowski-Sachs universe
\begin{equation}
ds^2=-dt^2+A^2dx^2+B^2(dy^2+dz^2)\label{eq:5}
\end{equation}
and a cosmic fluid with perfect fluid distribution $T_{\mu\nu}=(\rho+p)u_{\mu}u_{\nu}+pg_{\mu\nu}$, the field equations in the GBD theory are expressed as 

\begin{eqnarray}
2H_{x}H_{y}+H_{y}^{2} &=& \frac{\rho}{\phi}+\frac{\omega(\phi)}{2}\left(\frac{\dot{\phi}}{\phi}\right)^{2}
-3H\left(\frac{\dot{\phi}}{\phi}\right)+\frac{V(\phi)}{2\phi},\label{eq:6}\\
2\dot{H_{y}}+3H_{y}^{2} &=& \frac{-p}{\phi}-\frac{\omega(\phi)}{2}\left(\frac{\dot{\phi}}{\phi}\right)^{2}
-\frac{6H}{k+2}\left(\frac{\dot{\phi}}{\phi}\right)-\dfrac{\ddot{\phi}}{\phi}+\dfrac{V(\phi)}{2\phi}, \label{eq:7}\\
H_{x}^{2}+H_{y}^{2}+H_{x}H_{y}+\dot{H_{x}}+\dot{H_{y}} &=& -\frac{p}{\phi}-\frac{\omega(\phi)}{2}\left(\frac{\dot{\phi}}{\phi}\right)^{2}
-\dfrac{3(k+1)H}{k+2}\left(\frac{\dot{\phi}}{\phi}\right)-\dfrac{\ddot{\phi}}{\phi}+\dfrac{V(\phi)}{2\phi}.\label{eq:8}
\end{eqnarray}
Here $\rho$ is the dark energy density  and $p$ is the dark energy pressure. $A=A(t)$ and $B=B(t)$ are the time dependent metric potentials. The overhead dots over a field variable denote ordinary time derivatives. The directional Hubble parameters are derived from the metric potentials as $H_x=\frac{\dot{A}}{A}$ and $H_y=H_z=\frac{\dot{B}}{B}$.  Assuming a simple anisotropic relation among the directional Hubble parameters $H_{x}=kH_{y}$, we get the mean Hubble parameter as $H=\frac{1}{3}(k+2)H_{y}=\frac{1}{\xi}H_y$. Here we have defined an anisotropic parameter $\xi=\frac{3}{k+2}$. For an isotropic universe, $\xi=1$. If $k>1$, we have $\xi<1$ and on the other hand of $k<1$, $\xi$ will be less than $1$. In terms of the mean Hubble parameter $H$, the GBD field equations can be expressed as 
\begin{eqnarray}
(2k+1)\xi^2H^{2} &=&\frac{\rho}{\phi}+\frac{\omega(\phi)}{2}\left(\frac{\dot{\phi}}{\phi}\right)^{2}
-3H\left(\frac{\dot{\phi}}{\phi}\right)+\frac{V(\phi)}{2\phi}, \label{eq:9}\\
2\xi\dot{H}+3\xi^2H^{2} &=& \frac{-p}{\phi}-\frac{\omega(\phi)}{2}\left(\frac{\dot{\phi}}{\phi}\right)^{2}
-2\xi H\left(\frac{\dot{\phi}}{\phi}\right)-\dfrac{\ddot{\phi}}{\phi}+\dfrac{V(\phi)}{2\phi}, \label{eq:10}\\
(k+1)\xi\dot{H}+(k^{2}+k+1)\xi^2H^{2}&=& -\frac{p}{\phi}-\frac{\omega(\phi)}{2}\left(\frac{\dot{\phi}}{\phi}\right)^{2}
-(k+1)\xi H\left(\frac{\dot{\phi}}{\phi}\right)
-\dfrac{\ddot{\phi}}{\phi}+\dfrac{V(\phi)}{2\phi}.\label{eq:11}
\end{eqnarray}

The Klein-Gordon wave equation for the scalar field becomes
\begin{equation}
\dfrac{\ddot{\phi}}{\phi}+3H\frac{\dot{\phi}}{\phi}=\dfrac{\rho-3\rho}{2\omega(\phi)+3}-\dfrac{\frac{\partial\omega(\phi)}{\partial\phi}\dot{\phi}^{2}}{2\omega(\phi)+3}-\dfrac{2V(\phi)-\phi\frac{\partial V(\phi)}{\partial\phi}}{2\omega(\phi)+3}.\label{eq:12}
\end{equation}

The evolution equation for the BD scalar field are obtained from Eqs.\eqref{eq:10} and \eqref{eq:11} as
\begin{equation}
-\frac{\dot{H}}{H}-3H=\frac{\dot{\phi}}{\phi}.\label{eq:13}
\end{equation}
In terms of deceleration parameter (DP) $q=-1-\frac{\dot{H}}{H^2}$, this evolution equation can be expressed as

\begin{equation}
(q-2)H=\frac{\dot{\phi}}{\phi}.\label{eq:14}
\end{equation}

It should be mentioned here that a positive DP describes a decelerating universe whereas a negative $q$ implies an accelerating one. For a constant DP, Eq. \eqref{eq:14} can be integrated to obtain the BD scalar field  as 
\begin{equation}
\phi\simeq a^{n}, \label{eq:15}
\end{equation}
where $a$ is the radius scale factor and $n=q-2$. This result shows that a constant deceleration parameter favours a power law for the Brans-Dicke scalar field. It is not new that, power law scalar fields have been used in literature to address different issues in cosmology. If we assume that $q$ evolves from  positive values to negative values than for $q > 2$, the BD scalar field $\phi$ can be an increasing function of the scale factor. For $q < 2$, the exponent $n$ becomes a negative quantity and $\phi$ becomes a  decreasing function of the scale factor. In an accelerating universe, the DP has a negative value. Recent estimates predict a narrow range of DP as $-0.8\leq q \leq -0.4$. In view of these recent estimates, the BD scalar field becomes a decreasing function of the scale factor. 

The Brans-Dicke parameter and the self-interacting potential are obtained from the field equations  \eqref{eq:9}-\eqref{eq:11} as \cite{SKT15, SKT20}
\begin{eqnarray}
\omega(\phi) &=& \left(\frac{\dot{\phi}}{\phi}\right)^{-2}\left[-\dfrac{\rho+p}{\phi}-\frac{\ddot{\phi}}{\phi}+k\xi H\frac{\dot{\phi}}{\phi}-2\xi\dot{H}+2(k-1)\xi^2H^{2}\right],\label{eq:16}
\\V(\phi) &=& 2\phi\left[(2k+1)\xi^2 H^{2}-\frac{\rho}{\phi}-\frac{\omega(\phi)}{2}\left(\frac{\dot{\phi}}{\phi}\right)^{2}+3H\frac{\dot{\phi}}{\phi}\right].\label{eq:17}
\end{eqnarray}

The dynamical BD parameter and the self interacting potential depend on a dynamical relationship between the dark energy density and the pressure and a presumed dynamics (known behaviour of DP) of the universe. The scale factor of the universe can be fixed from the behaviour of the deceleration parameter. We have already mentioned earlier that, a constant DP leads to a power law behaviour of the BD scalar field. A constant DP can be obtained from a power law expansion or a de Sitter kind of expansion of the scale factor. However in the present work, we wish to explore the possibility of a time varying DP with a signature flipping behaviour. Such a dynamical DP obviously changes the nature of the BD scalar field with a more involved expression for it. In order to obtain the dynamical BD parameter and the self interacting potential, we consider here a simple linear equation of state which is known as the equation of state of unified dark fluid (UDF). The UDF has two different aspects that explain the unification of both the dark sectors: dark energy and dark matter. It is worth to mention here that, there are a good number of generalised equation of states available in literature where the dark energy and dark matter are considered to be two different aspects of the same cosmic fluid \cite{Capo2006}. In these generalized models, the equation of state depends not only on the energy density but also on the Hubble parameter and its derivatives.

\section{Unified Dark Fluid}

It is certain that, the late time cosmic acceleration is mostly triggered by the presence of exotic dark energy and non baryonic matter. Estimate as per the recent Planck data shows that the dark sector comprising dark energy and dark matter has a lion share of around $95\%$ in the mass-energy budget of universe i.e. $68.3\%$ dark energy, $26.8\%$ dark matter and $4.9\%$ baryonic matter. There are many explanation for the bizarre late time cosmic speed up phenomenon. Different concepts and ideas have been developed to understand the phenomena. A specific approach that unifies the dark matter and dark energy into a single UDF may be useful in this direction. A dark fluid model with a linear equation of state 
\begin{equation}
p=\alpha(\rho-\rho_{0}),\label{eq:18}
\end{equation}
was proposed in the spirit of generalized Chaplygin gas model(CGM) after its success in addressing issues related to the late time cosmic acceleration and dark energy problem \cite{Babichev2005}. Here $\alpha$ and $\rho_{0}$ are constant parameters of UDF. This non-homogeneous linear equation of state provides a description of both hydro-dynamically stable $(\alpha > 0)$ and unstable $(\alpha< 0)$ fluids. The UDF contains two parts. One behaves as a barotropic cosmic fluid and the other behaves as a cosmological constant. This is usually referred to as dark degeneracy. The UDF has a constant adiabatic speed of sound, $C_{s}^{2}=\frac{dp}{d\rho}=\alpha$. The dark matter sector is represented by $\alpha=0$.  $\alpha=1$ implies a stiff fluid dominated with dark energy. An intermediate value of $\alpha$ refers to an exotic cosmic fluid unifying the dark sector of the universe. In some recent works, the value of the adiabatic sound speed $\alpha$ for the UDF  has been constrained to be $\alpha=0.000487_{-0.000487}^{+0.000787}$ \cite{Xu12} and $\alpha=0.00172_{-0.00497}^{+0.00392}$ \cite{Liao12}. Holeman and Naidu in their work \cite{Holeman2005}, claimed that the UDF model is consistent with SNIa observations \cite{A.G 2004}, WMAP \cite{D.N 2007} and constraints from matter power spectrum measurements \cite{Tegmark02}. They have constrained the adiabatic sound speed parameter as $\alpha=0.316$. In the present work, we will use these constrained values of the adiabatic sound speed to study the late time dynamics in the frame work of GBD theory.

Integrating the conservation equation
\begin{equation}
\dot{\rho}+3H\left({p}+{\rho}\right)=0,\label{eq:19}
\end{equation}
using the UDF linear equation of state Eq.\eqref{eq:18}, the dark energy density can be obtained as
\begin{equation}
{\rho}={\rho_{\Lambda}}+{\rho_{\alpha}}(a^{-3(1+\alpha)}).\label{eq:20}
\end{equation}
Consequently, the dark energy pressure becomes
\begin{equation}
p={-\rho_{\Lambda}}+{\alpha}{\rho_{\alpha}}(a^{-3(1+\alpha)}),\label{eq:21}
\end{equation}
where $\rho_{\Lambda}=\dfrac{\alpha\rho_{0}}{1+\alpha}$ and $\rho_{\alpha}=\rho_{0}-\rho_{\Lambda}$. $\rho_{0}$ is the dark energy density at the present epoch.

The dark energy equation of state $\omega_{D}=\frac{p}{\rho}$ can  be obtained from the Eqs.\eqref{eq:20} and \eqref{eq:21} as
\begin{equation}
\omega_{D}=-1+\dfrac{1+\alpha}{1+\left(\dfrac{\rho_{\Lambda}}{\rho_{\alpha}}\right)(a)^{3(1+\alpha)}}.\label{eq:22}
\end{equation}

In terms of the redshift, the dark energy equation of state can be expressed as
\begin{equation}
\omega_{D}=-1+\dfrac{1+\alpha}{1+\left(\dfrac{\rho_{\Lambda}}{\rho_{\alpha}}\right)(1+z)^{-3(1+\alpha)}},\label{eq:23}
\end{equation}
where we have used the fact $1+z=\frac{1}{a}$.

\begin{figure}[t]
\begin{center}
\includegraphics[width=0.8\textwidth]{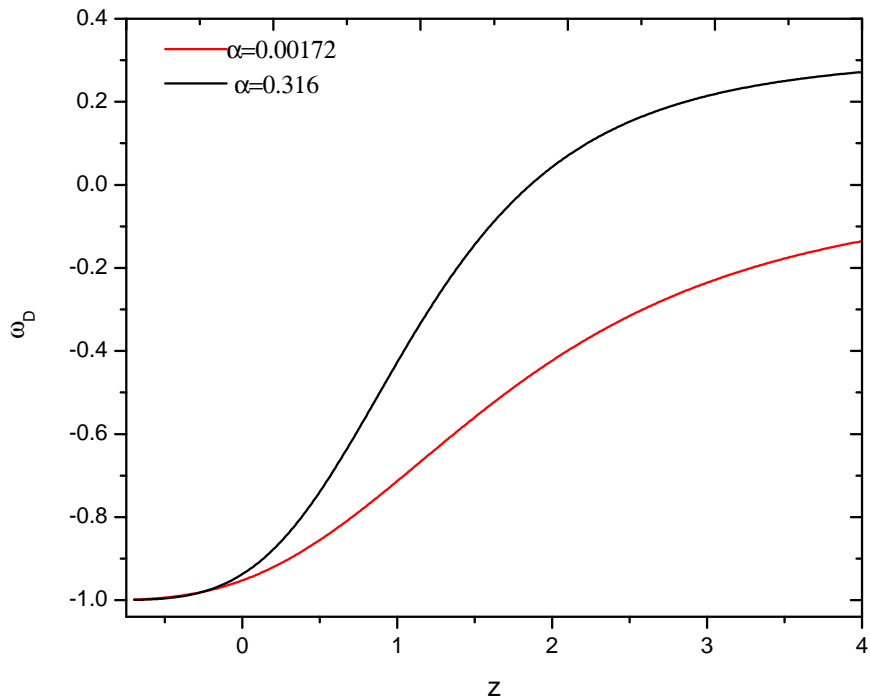}
\caption{Dark energy equation of state as a function of red shift for two different values of $\alpha$. The ratio $\frac{\rho_{\Lambda}}{\rho_{\alpha}}$ is taken to be 20.}
\end{center}
\end{figure}
In Figure 1, we have shown the evolutionary behaviour of the dark energy equation of state for two different values of the adiabatic sound speed as has been constrained in some recent works. The dynamical behaviour of the dark energy equation of state for UDF have been discussed in the context of a bouncing model using different values of the ratio $\dfrac{\rho_{\Lambda}}{\rho_{\alpha}}$ in Ref. \cite{SKT20}. As has been observed in that work, a change in the value of this ratio does not change the overall dynamical behaviour of $\omega_D$. In view of this, in the present work, we have considered $\dfrac{\rho_{\Lambda}}{\rho_{\alpha}}=20$ to plot the figure. The lower curve in the figure corresponds to low value of $\alpha$ ($\alpha=0.00172$) where as the upper curve is for the higher value ($\alpha=0.316$). Different values of $\alpha$ lead to different evolution path history of the dark energy equation of state. In other words, $\alpha$ affects the slope of $\omega_{D}$ in the sense that, higher the value of $\alpha$ higher is the slope. However at late times,  $\omega_{D}$ evolves to overlap with $\Lambda$CDM model ($\omega_{D}=-1$) irrespective of the value of $\alpha$.

\begin{figure}
\centering
\includegraphics[width=0.7\textwidth]{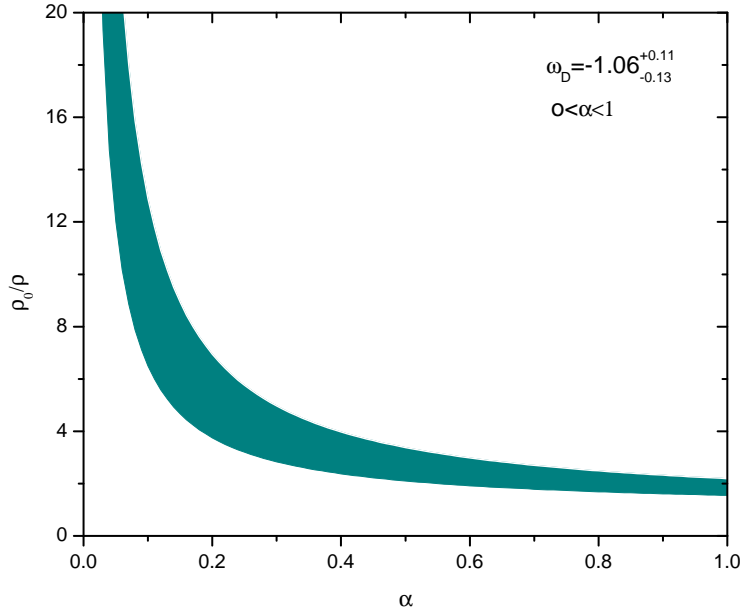}
\caption{Observational constraints on the parameters of unified dark fluid. The dark energy equation of state is taken from Ref.\cite{Kumar14}. The square of adiabatic sound speed is considered to be positive corresponding to non phantom models. This figure is similar to the Fig.1 of Ref. \cite{SKT15}. } 
\end{figure}

\begin{figure}
\includegraphics[width=0.7\textwidth]{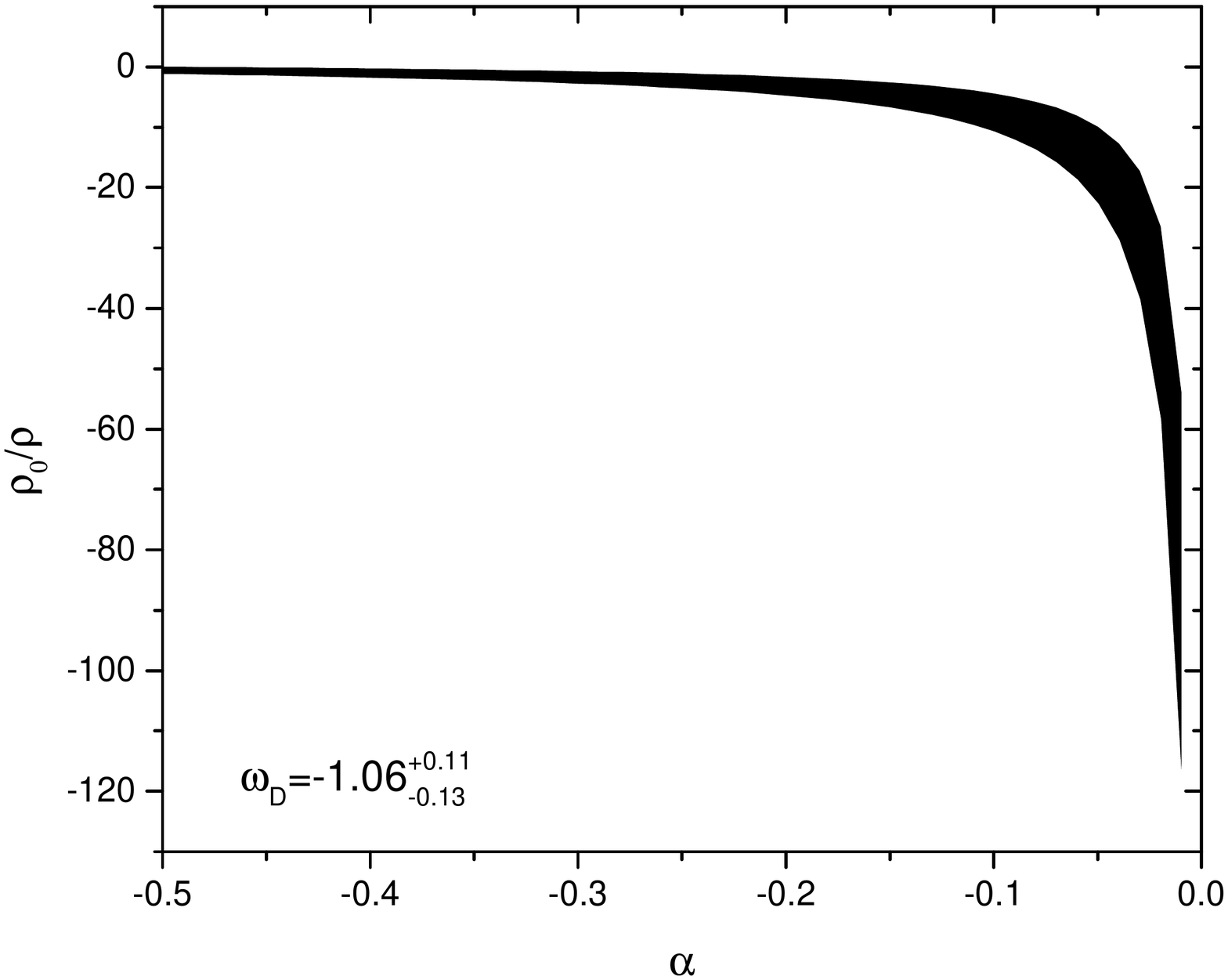}
\caption{Observational constraints on the parameters of unified dark fluid. The dark energy equation of state is taken from Ref.\cite{Kumar14}. The square of adiabatic sound speed is considered to be negative corresponding to phantom models.}
\end{figure}

\subsection{Observational constraints on model parameters}

The parameters of UDF can be constrained from different observational and theoretical aspects. In the present work, we have used the constrained values of the dark energy equation of state parameter to constrain the adiabatic sound speed $\alpha$ and $\rho_0$. Since $\alpha$ can be positive for non-phantom models and negative for phantom models, we have tried to put some observational constraints on the model parameters separately for the  non-phantom and phantom cases. In a recent work, Tripathy et al. \cite{SKT15} have constrained the UDF model parameters in the non phantom region only. The observational constraint for the model parameter $\rho_{0}$ normalised to dark energy density is shown in Figure 2 for the physically allowed range $0<\alpha<1$ for non-phantom models. In this figure we have considered a constant dark energy equation of state $\omega_{D}=-1.06_{-0.13}^{+0.11}$ as constrained in a recent work\cite{Kumar14}. This figure is the same as that of Ref. \cite{SKT15}. The lower boundary in the figure corresponds to the lower value of $\omega_{D}$ and upper boundary corresponds to upper value of $\omega_{D}$. The shaded region in between the lower and upper boundary constrain the model parameters. In general, $\dfrac{\rho_{0}}{\rho}$  decreases with the increase in $\alpha$. It is evident from the plot that for non-phantom models with alpha lying in the range $0<\alpha<1$, $\rho_{0}$ is constrained to be positive. One can also note that for small values of $\alpha$ i.e $0<\alpha<0.1$, $\dfrac{\rho_{0}}{\rho}$ is much larger with a narrow uncertainty range. Similarly for values of alpha closer to 1, $\dfrac{\rho_{0}}{\rho}$ can be constrained within a narrow range but with smaller values. However for $\alpha$ lying in the range $0.1<\alpha<0.5$ a much wider range of  $\dfrac{\rho_{0}}{\rho}$ is allowed.  


For phantom models $\alpha$ can take negative values. We have tried to constrain the values of $\dfrac{\rho_{0}}{\rho}$ for negative values of $\alpha$ in Figure 3 with $\omega_{D}=-1.06_{-0.13}^{+0.11}$. The lower boundary in the figure corresponds to the lower value of $\omega_{D}$ and upper boundary corresponds to the upper value of $\omega_{D}$. The shaded region in between the lower and upper boundary constrain the model parameters. From this figure, it is clear that for a decrease in the value of $\alpha$, $\dfrac{\rho_{0}}{\rho}$ increases from larger negative values to zero. For $\alpha>-0.05$  and $\alpha<-0.03$ the parameters are constrained to a very narrow range. T

\begin{figure}[h!]
\begin{center}
\includegraphics[width=0.65\textwidth]{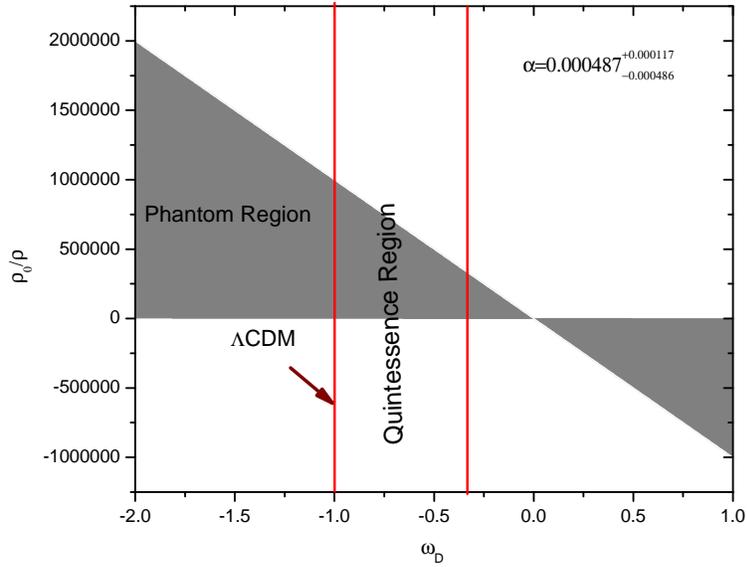}
\caption{Observational constraints on the parameters of unified dark fluid. The model parameter $\rho_0$ is plotted as a function of dark energy equation of state with a fixed value of $\alpha=0.000487_{-0.000487}^{+0.000787}$ taken from Ref.\cite{Xu12}.  The shaded region shows the constrained values of the parameter.}
\end{center}
\end{figure}
\begin{figure}[h!]
\begin{center}
\includegraphics[width=0.65\textwidth]{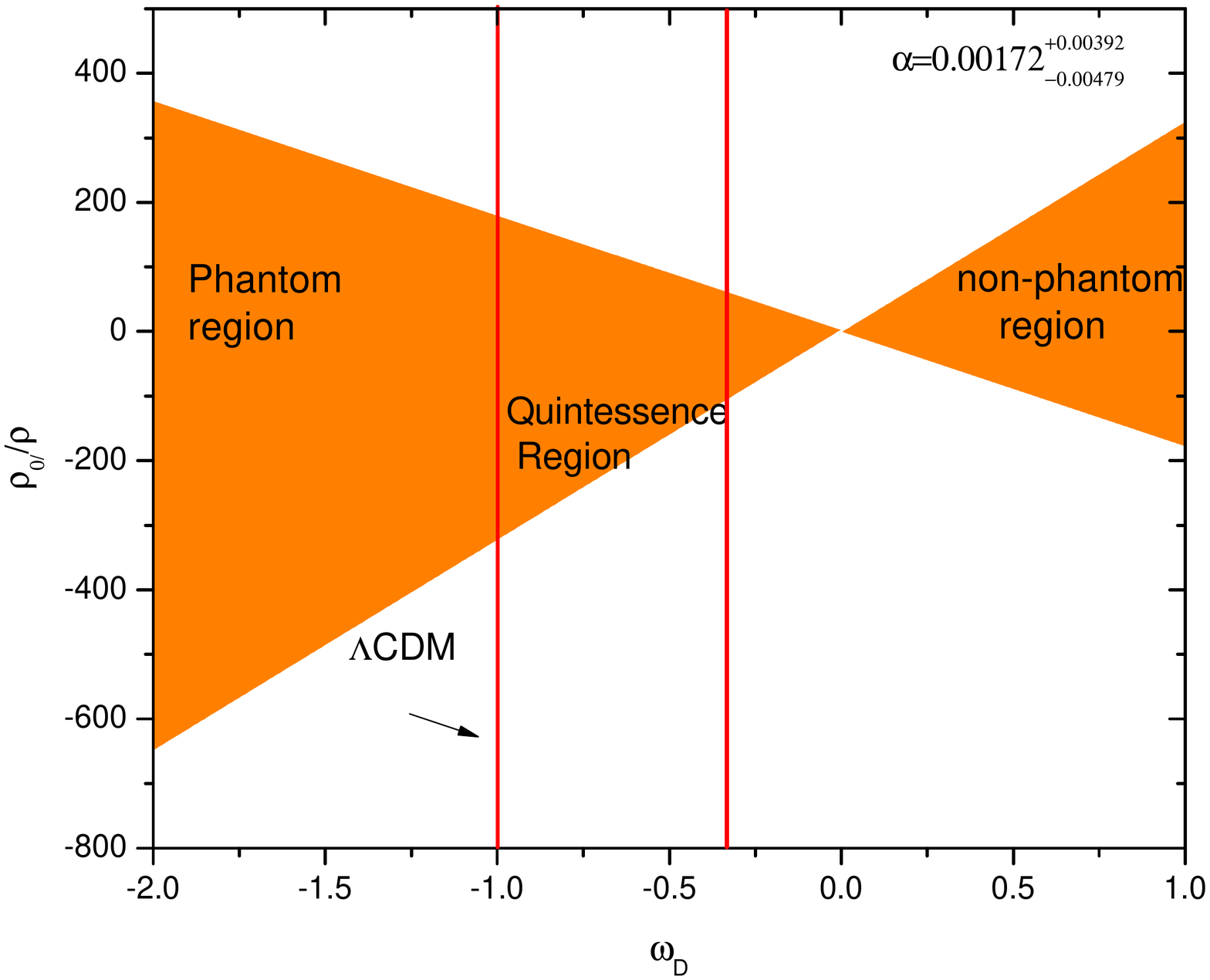}
\caption{Observational constraints on the parameters of unified dark fluid. The model parameter $\rho_0$ is plotted as a function of dark energy equation of state with fixed value of $\alpha=0.00172_{-0.00497}^{+0.00392}$ taken from  Ref.\cite{Liao12}. The shaded region shows the constrained values of the parameter.}
\end{center}
\end{figure}
he model parameter $\rho_{0}$ can also be constrained basing upon a variation of dark energy equation of state with a constrained values of $\alpha$. In some recent works $\alpha$ has been constrained to be $\alpha=0.000487_{-0.000487}^{+0.000787}$ \cite{Xu12} and
$\alpha=0.00172_{-0.00497}^{+0.00392}$ \cite{Liao12}. In Figures 4 and 5, we have shown the observational constraints on the model parameters by plotting $\dfrac{\rho_{0}}{\rho}$ as the function of $\omega_{D}$. In these figures, two values of $\alpha$ are considered. In both the figures, the dark energy equation of state $\omega_{D}$ has been varied from some phantom regions to a stiff fluid through quintessence region and dust universe ($\omega_{D}=0$). One interesting aspect of these constraints is that, for $\alpha=0.000487_{-0.000487}^{+0.000787}$, the constrained values of $\frac{\rho_0}{\rho}$ lie in the positive domain for phantom region and quintessence region and have negative values for dust and matter dominated non-phantom region. However, for $\alpha=0.00172_{-0.00497}^{+0.00392}$, in both the phantom and non-phantom regions, the constrained values of $\frac{\rho_0}{\rho}$ can be both positive and negative depending upon the fixed dark energy equation of state parameter. In both the figures one can note that the parameters can be constrained to a narrow band in the quintessence region compared to that in phantom region.

\subsection{Unified Dark Fluid in $\omega^{\prime}-\omega$ plane}

From the conservation equation Eq. \eqref{eq:19}, we get 
\begin{equation}
\omega^{\prime}=-3(1+\omega)(\dfrac{dp}{d\rho}-\omega),\label{eq:24}
\end{equation}
where $\omega^{\prime}=\dfrac{d\omega}{d(lna)}$. For the UDF equation of state, the above equation reduces to
\begin{equation}
\omega^{\prime}=-3(1+\omega)(\alpha-\omega).\label{eq:25}
\end{equation}

\begin{figure}[h!]
\begin{center}
\includegraphics[width=0.65\textwidth]{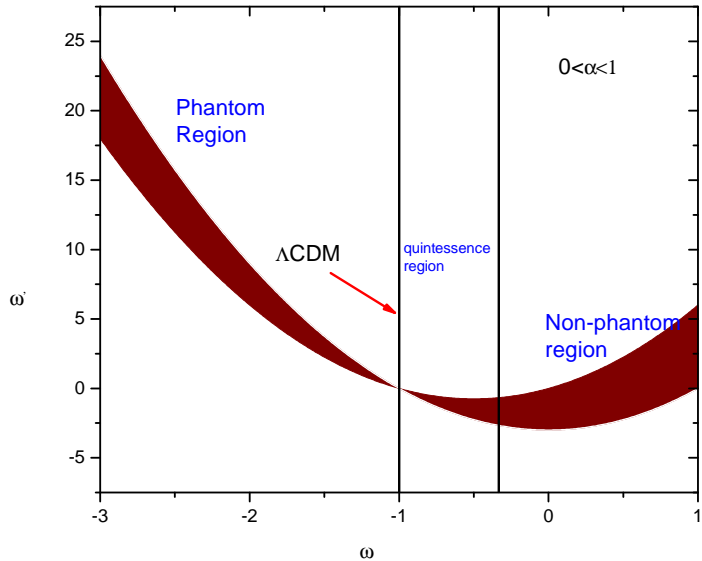}
\caption{Behaviour of Unified dark fluid in $\omega^{'}-\omega$} plane
\end{center}
\end{figure}
 
In Figure 6, we have shown the allowed boundaries of the UDF in the $\omega^{\prime}-\omega$ plane for both phantom and non-phantom models. The UDF resembles thawing models in the quintessence region. Unified dark fluid with constrained values of $\alpha$ as in Refs. \cite{Liao12, Xu12} overlaps with the upper boundary in the non-phantom region and with lower boundary in the phantom region.

\section{UDF Model with a signature flipping deceleration parameter}

Observations have confirmed that the universe is having an accelerated expansion at the present epoch. Also, it is believed that the universe may have undergone a transition from a decelerated phase to an accelerated one. The cosmic redshift at which this transition occurs is called transition redshift $z_{da}$. This implies a signature flipping behaviour of the deceleration parameter $q$ from a positive value at some early epoch to negative values at late times of cosmic evolution. In the present work, we wish to construct some cosmological models that are consistent with this behaviour of the universe and can provide a signature flipping of the deceleration parameter. In view of this, we consider a hybrid scale factor, $a=a_0\left(\frac{t}{t_{0}}\right)^{H^{1}}e^{H^{0}(t-t_{0})}$ where $H^{0}$ and $H^{1}$ are positive constants and $a_0$ is the scale factor at the present epoch $t_0$. In the present work, we have assumed $a_0=t_0=1$. This hybrid scale factor has two factors: one factor behaving like exponential expansion and the other factor behaving like power law expansion. While the power law behaviour dominates the cosmic dynamics in early phase of cosmic evolution, the exponential factor dominates at late phase. When $H^{1} = 0$, the exponential law is recovered and for $H^{0} = 0$, the scale factor simulates a power law expansion. In an earlier work, Tripathy et al. \cite{SKT15} have used power law and exponential law of expansion to construct some UDF models in the framework of GBD theory. The power law and the exponential law of expansion invokes a constant deceleration parameter. Keeping in view of the cosmic transit behaviour, we wish to use the hybrid expansion law where the deceleration parameter shows a transition from  positive values at an early epoch to negative values at late time of cosmic evolution. The Hubble parameter for this model is obtained as $H=H^{0}+\frac{H^{1}}{t}$ and the deceleration parameter as $q=-1+\frac{H^{1}}{(H^{0}t+H^{1})^{2}}$. The HSF has already been considered in some recent works by Mishra and Tripathy in the framework of GR and modified gravity to address different issues concerning the late time cosmic speed up phenomena \cite{Mishra15, Mishra2018, Mishra2018a, Ray2019, Mishra2019, Mishra19a}. In a recent work, Tripathy \cite{SKT2014} has considered a more general form of such hybrid Hubble parameter  with the form $H=H^{0}+\frac{H^{1}}{t^{m}}$ where $m$ is a positive constant. The present HSF is a special case of that considered in Ref.\cite{SKT2014} with $m = 1$. The deceleration parameter for the HSF reduces to $q\simeq-1+\frac{1}{H^{1}}$ as $t \longrightarrow 0$ and evolves with cosmic time to become $q\simeq-1$ at late phase of cosmic evolution as $t \longrightarrow\infty$. The transit epoch, when the universe flips from a decelerated phase to an accelerated one corresponds to a redshift called transit redshift $z_{da}$. Different observations  and theoretical considerations have constrained this parameter to be an order of 1 i.e $z_{da} \sim 1$. For example, Busca \cite{Busca13} constrained the transition redshift as $z_{da}= 0.82\pm 0.08$;  Farooq and Ratra constrained this as $z_{da}=0.74\pm 0.05$ \cite{Farooq13}, Capozziello et al. obtained a constraint on this parameter as $z_{da}= 0.7679^{+0.1831}_{-0.1829}$ \cite{Capo14}.  Reiss et al. derived  kinematic limits on the transition redshift as $z_{da}= 0.426^{+0.27}_{-0.089}$ \cite{Reiss07}. While Lu et al. constrained this parameter to be $z_{da}=0.69^{+0.23}_{-0.12}$ \cite{Lu11}, Moresco et al. obtained the constraint $z_{da}=0.4\pm 0.1$ \cite{Moresco16}. At the transit redshift, the deceleration parameter vanishes and this corresponds to a cosmic time $t=-\frac{H^1}{H^0}\pm \frac{\sqrt{H^1}}{H^0}$. In the context of standard Big Bang cosmology with positive time frame only, one may have the transition epoch as $t=\frac{\sqrt{H^1}-H^1}{H^0}$. Obviously, it constrains the parameter $H^1$ in the range $0<H^1<1$. In a recent work, Mishra and Tripathy have tried to constrain this parameter in the range $0<H^1<\frac{1}{3}$ \cite{Mishra15}. In that work, $H^0$ is taken as a free parameter. However, in a subsequent work, $H^0$ is constrained in the range $0.075 \leq H^0 \leq 0.1$ according to the constraints on the transition redshift $0.4\leq z_{da}\leq 0.8$ \cite{Ray2019}. Further, Mishra et al. used the specific values $H^0=0.695$ and $H^1=0.085$ to reproduce the transition redshift of $z_{da}=0.806$ \cite{Mishra2018}. In that work, Mishra et al. have shown that, the dark energy equation of state obtained in the HSF model in the framework of an extended gravity is quite comparable with the parametrized equation of states such the CPL \cite{Cheval2001, Linder2003} and BA \cite{Barboza2012}. It is certain that, HSF is required to simulate the signature flipping behaviour of the deceleration parameter. Out of the two parameters of HSF, the value of $H^0$ decides the rate of transition from a decelerated universe to an accelerated one. Higher the value $H^0$, faster is the rate of transition. Therefore, these parameters can be suitably constrained from the observational transit redshift values as well as the Hubble parameter data at different redshifts. With a motivation to constrain the parameters of HSF, in a recent work, we have considered two specific values of $H^1$ namely $0.2$ and $0.3$ and constrained the other parameter to get two specific values of the transition redshift i.e. $0.8$ and $0.5$ \cite{SKT2020}. In this process, we have obtained four different models namely HSF11, HSF21, HSF12 and HSF22 and compared the constructed models with the observational Hubble parameter data.. The model parameters as obtained are given in Table I. The constructed models are well within the acceptable range of the observed value. Also the constructed models which are stiff at higher redshift (fast deceleration) are observed to be soft at low redshift (slow acceleration). The predicted values of different properties of the universe at the present epoch from the constructed models are given in Table II. The predicted values of the deceleration parameter from the models lie in between $-0.494$ and $-0.723$ and are close to the estimates from recent data. We have adopted the same comparison figure from the Ref. \cite{SKT2020} for clarity.

 \begin{figure}[t]
 \begin{center}
 \includegraphics[width=0.7\textwidth]{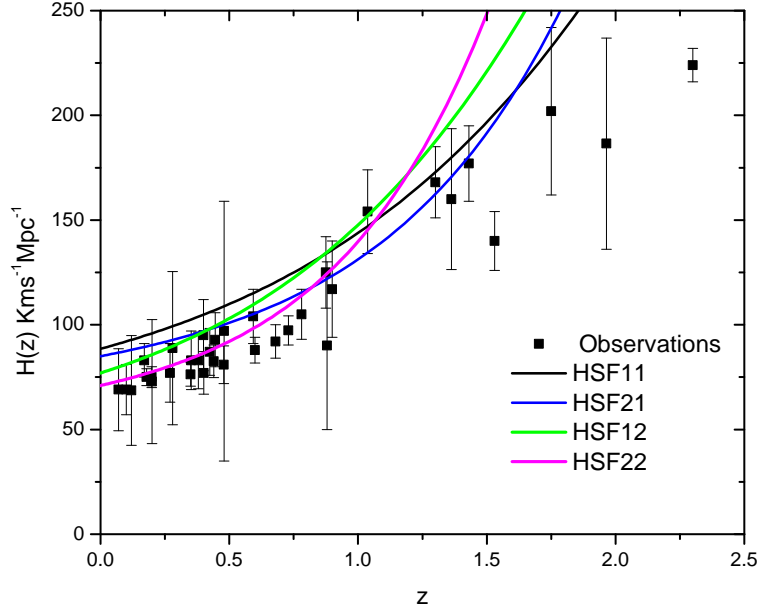}
 \caption{ Hubble parameter $H(z)$ as function of redshift $z$. The unconnected solid squares with error bars are the observational data for Hubble parameter. The constructed models HSF11, HSF21, HSF12 and HSF22 are compared with the observational data. The figure is taken from Ref. \cite{SKT2020}.}
 \end{center}
 \end{figure}

\begin{table}
\caption{Model parameters of the hybrid scale factor as constrained from transition redshift data \cite{SKT2020}.}
\centering
\begin{tabular}{c|c|c|c}
\hline
\hline
Models & $H^1$& $H^0$& $z_{da}$\\
\hline
HSF11&0.3&0.585&0.8\\
HSF21&0.2&0.65&0.8\\
HSF12&0.3&0.47&0.5\\
HSF22&0.2&0.51&0.5\\
\hline
\end{tabular}
\end{table}

\begin{table}
\caption{ Predicted values of Hubble rate, deceleration parameters, Brans-Dicke parameter,  Self interacting potential and $\frac{\dot{G}}{G}$ at the present epoch.}
\centering
\begin{tabular}{c|c|c|c|c|c}
\hline
\hline
Models & $H_0$& $q_0$& $\omega _0$& $V(\phi_0)$&$\left(\frac{\dot{G}}{G}\right)_0$ ($\times 10^{-10} yr^{-1}$) \\
\hline
HSF11&88.5&-0.6169&-1.3349&-0.858&1.9096$\leq \left(\frac{\dot{G}}{G}\right)_0 \leq 1.9938$\\
HSF21&85&-0.723&-1.328&-0.769&1.9098$\leq \left(\frac{\dot{G}}{G}\right)_0 \leq 1.9931$\\
HSF12&77&-0.494&-1.314&-0.8856&1.7825$\leq \left(\frac{\dot{G}}{G}\right)_0 \leq 1.8610$\\
HSF22&71&-0.6032&-1.3039&-0.814&1.7699$\leq \left(\frac{\dot{G}}{G}\right)_0 \leq 1.8481$\\
\hline
\end{tabular}
\end{table}

Once the DE equation of state is fixed and the models are constructed basing upon certain physical basis, it becomes straightforward to study the evolutionary aspect of the BD scalar field, BD parameter, self interacting potential and the Newtonian gravitational constant. In the following subsections we have investigated the behaviour of these quantities for the four HSF models taken along with the UDF equation of state.

\begin{figure}[t]
\begin{center}
\includegraphics[width=0.7\textwidth]{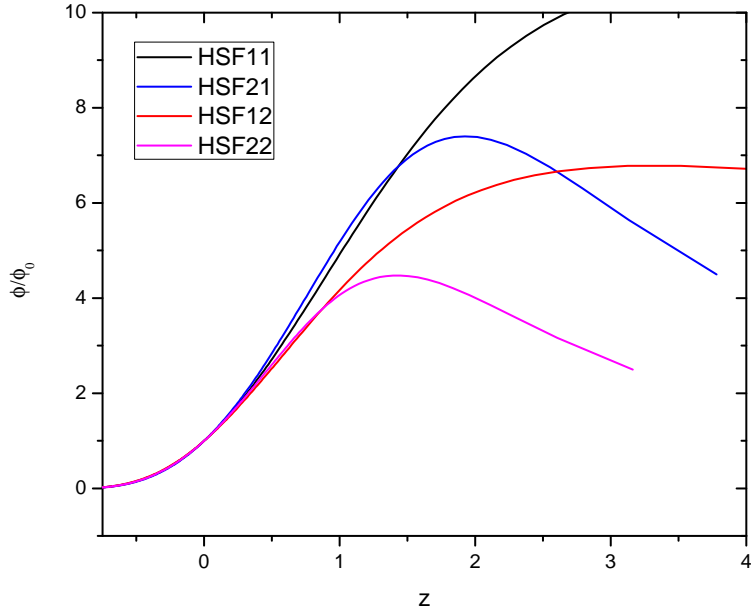}
\caption{ Evolution of the Brans-Dicke scalar field $\phi$.}
\end{center}
\end{figure}

\subsection{Brans-Dicke scalar field}
The evolution equation for the Brans-Dicke scalar field is given by $\frac{\dot{\phi}}{\phi}=-\frac{\dot{H}}{H}-3H$ which on integration for an HSF yields
\begin{equation}
\phi= \phi_{0}\left(H^{0}+H^{1}\right) \frac{a^{-3}}{H} ,\label{eq:26}
\end{equation}
where, $\phi_{0}$ is the value of the BD scalar field at the present epoch. The BD scalar field depends on the scale factor and its first derivative.  From the above expression of the BD scalar field, we can recover the results of Ref. \cite{SKT15} for an exponential and power law expansion of the scale factor. For an exponential expansion of the scale factor, the Hubble rate becomes a constant and consequently the BD scalar field reduces to  $\phi=\phi_0a^{-3}$. Similarly for a power law expansion such as $a\sim t^{H^1}$, the BD scalar field becomes $\phi \sim t^{\left(1-3H^1\right)}$. The BD scalar field can be expressed in terms of the redshift as $\phi = \phi_{0}\left(H^{0}+H^{1}\right) \frac{\left(1+z\right)^3}{H(z)}$, where $H(z)= \frac{1}{1+z}\frac{dz}{dt}$. In Figure 5, the evolution of the BD scalar field is shown as a function of redshift for the constructed models. The BD scalar field, for all the constructed models, decreases from some large values at early phase to small values at the present epoch. However, for the models HSF21 and HSF22  with $H^1=0.2$, $\phi$ first increases at an initial phase to a certain maximum and than decreases to the usual behaviour at some low redshift. At low redshift, $\phi$ behaves as $\left(1+z\right)^3$ irrespective of the parameters of the HSF models. It is interesting to note that, even if $\phi$ behaves alike for all the HSF models in the present epoch and afterwards, they exhibit quite different behaviours at a past cosmic phase  for  redshift $z>0.5$. The behaviour of the models HSF11 and HSF12 concerning the evolution of the BD scalar field appears to be smooth. In view of this, these two models may be more favourable for cosmological studies than the other two models. 

\subsection{Brans-Dicke parameter}
In the GBD theory, the BD parameter $\omega$ is a dynamically varying quantity and depends on the dark energy equation of state $\omega_D$ and the BD scalar field. A general expression for $\omega$ has been derived in Ref. \cite{SKT15}. Recalling that the BD scalar field depends on the Hubble rate through the relation $-\frac{\dot{H}}{H}-3H=\frac{\dot{\phi}}{\phi}$, the BD parameter can be expressed as
\begin{eqnarray}\label{omega}
\omega(\phi)&=&\left(\frac{\dot{H}}{H}+3H\right)^{-2} \\ \nonumber &\times& \left[-\dfrac{\rho+p}{\phi}+\frac{\ddot{H}}{H}-2\left(\frac{\dot{H}}{H}\right)^{2}-8\dot{H}-\left\{9+k\xi-2\left(k-1\right)\xi^2\right\}H^{2}\right],\label{eq:27}
\end{eqnarray}
where $\rho+p=(1+\alpha)\rho_{\alpha}a^{-3\left(1+\alpha\right)}$. In an earlier work \cite{SKT15}, UDF has been considered within the framework of GBD theory where the Brans-Dicke parameter $\omega$ has been calculated from the constructed anisotropic models assuming either a power law or an exponential expansion of the volume scale factor. Such assumption provides a constant deceleration parameter. In those cases it has been observed that, the anisotropic parameter affects only the non evolving part of the Bran-Dicke parameter. However, in the present work, we have considered a hybrid expansion law that mimics the real transient universe with early deceleration and late time cosmic acceleration. In this situation, it is clear from the expression of $\omega$ in Eq. \eqref{eq:27} that, the anisotropic parameter is more involved in the evolution of the BD parameter. 

\begin{figure}[t]
\begin{center}
\includegraphics[width=0.7\textwidth]{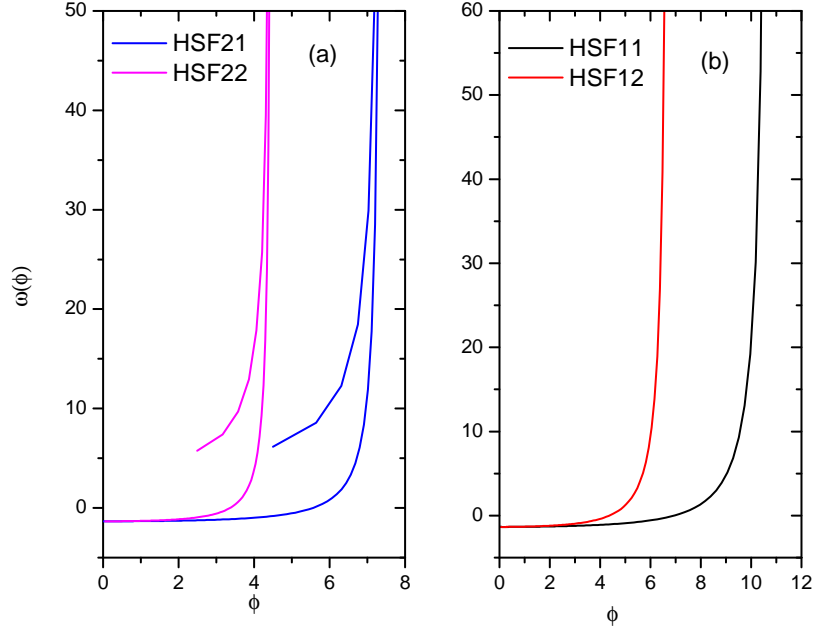}
\caption{ The Brans-Dicke parameter $\omega(\phi)$ as function of the scalar field $\phi$. In the left panel we have taken the models for $H^1=0.2$ and the right panel is for $H^1=0.3$. Here, the unified dark fluid parameter $\alpha$ is taken to be $0.00172$.}
\end{center}
\end{figure}

In Figures 9(a) and (b), we have shown the evolution of the Brans-Dicke parameter with respect to the BD scalar field. In Fig. 9(a), the BD parameter is shown for the two models with $H^1=0.2$ i.e. HSF21 and HSF22 and in the Fig. 9(b), it is shown for the models with $H^1=0.3$. In order to plot the figures, we have considered $\alpha=0.00172$. The BD parameter in general increases with the increase in the scalar field. However, for all the models, it is observed that, $\omega$ behaves alike at late times of cosmic evolution. For the cases with $H^1=0.2$, the behaviour of $\omega$ is peculiar at some early epochs forming a kind of loop structure with the scalar field. This behaviour may be due to the peculiarity  displayed by the BD scalar field for these models HSF21 and HSF22, where $\phi$ increases with $z$ and attains a maximum at certain $z$ and then decreases with the further increase in $z$. In view of this, the other two models, HSF11 and HSF12 may be more favourable for different studies in cosmology. It is interesting to note that, at a late phase of cosmic evolution, the BD parameter becomes a constant quantity and is independent of the choice of the HSF parameters. Regarding the quantitative estimate of the BD parameter $\omega$, there remains a good deal of controversy. For some cases it requires a very large value about 40,000. However, in the present work, we observed that all the models predict almost same value of $\omega$ close to $-1.3$ which is higher than the value $-1.5$ required for acceleration in isotropic models. In view of this, it can be assessed that, the anisotropy in the expansion rates has some role in deciding the value of $\omega$ at a given epoch.

\begin{figure}[h!]
\begin{center}
\includegraphics[width=0.7\textwidth]{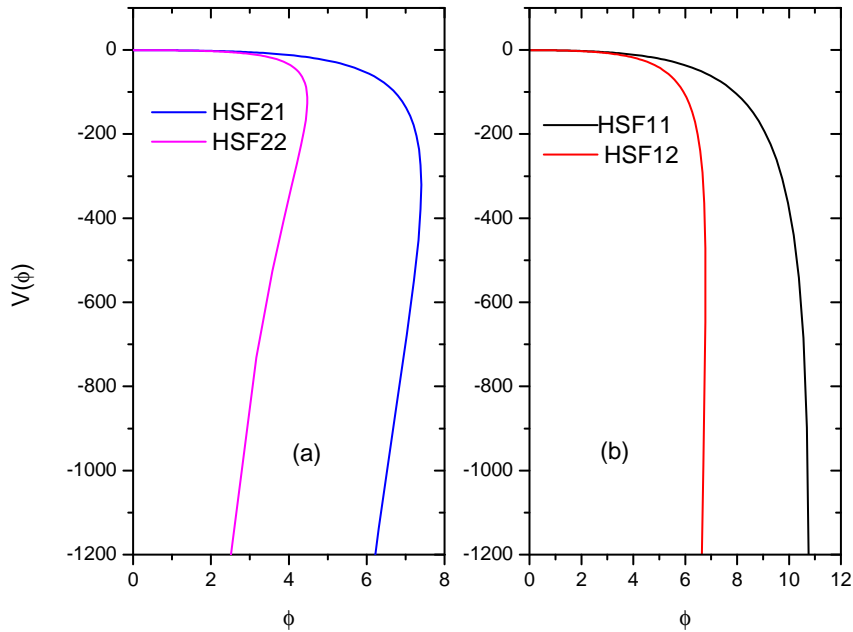}
\caption{ The self interacting potential $V(\phi)$ is shown as a function of scalar field $\phi$. In the left panel, $V(\phi)$ for models with  $H^1=0.2$ are  shown. The figures in the right panel are for $H^1=0.3$.}
\end{center}
\end{figure}

\subsection{Self interacting potential}
It is straightforward to calculate the self interacting potential $V(\phi)$ for the HSF models with UDF as
\begin{eqnarray}
V(\phi) &=& 2\phi\left[(2k+1)\xi^2 H^2-\frac{\rho_{\Lambda}+\rho_{\alpha}(a)^{-3(1+\alpha)}}{\phi}-\frac{\omega(\phi)}{2}\left(\frac{\dot{\phi}}{\phi}\right)^{2}+\frac{3\dot{\phi}}{\phi} H\right].
\end{eqnarray}
The time variation of the self interacting potential $V(\phi)$ is shown as function of the BD scalar field in Figures 10 (a) and (b). In the left panel of the Figure 10, $V(\phi)$ for models with  $H^1=0.2$ are  shown and the figures in the right panel are for $H^1=0.3$. Here also, we assume the UDF parameter $\alpha$ to be $0.00172$. The self interacting potential for the constructed HSF models are attractive in nature and increases sharply from some large negative value at an early epoch to vanishingly small values at late times of evolution. The four models behave alike in low redshift region where the scalar field assumes a smaller value. However, in the high redshift region with large scalar field, the four models splits to behave differently. At late phase of cosmic evolution, the self interacting potential becomes independent of the HSF models and appears to be non-evolving with respect to the BD scalar field. In fact, $V(\phi)$ asymptotically evolves to vanish at late times.

\begin{figure}[h!]
\begin{center}
\includegraphics[width=0.7\textwidth]{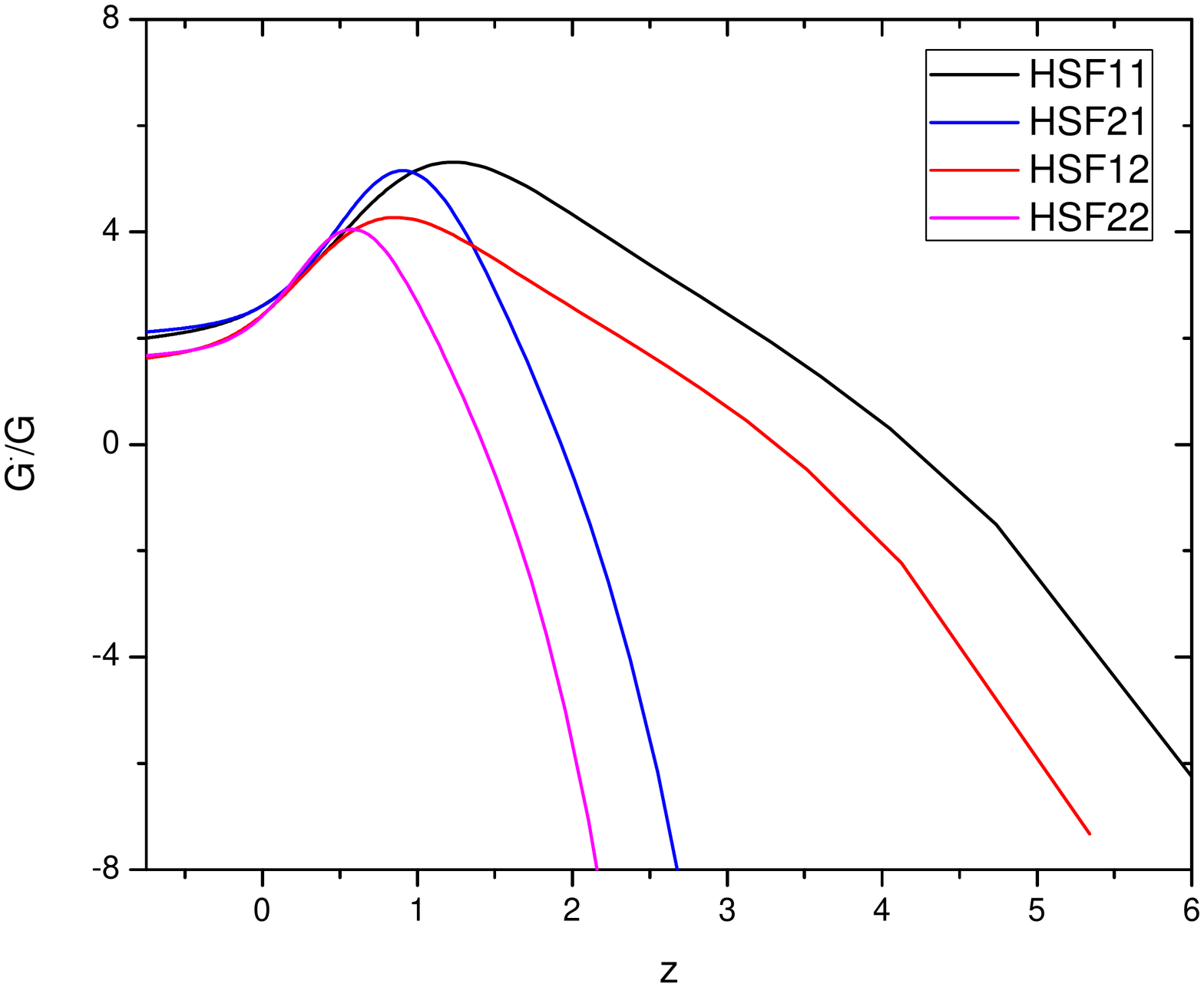}
\caption{ $\frac{\dot{G}}{G}$ as function of redshift $z$. The unified dark fluid parameter $\alpha$ is taken to be $0.00172$.}
\end{center}
\end{figure}

\begin{figure}[h!]
\begin{center}
\includegraphics[width=0.7\textwidth]{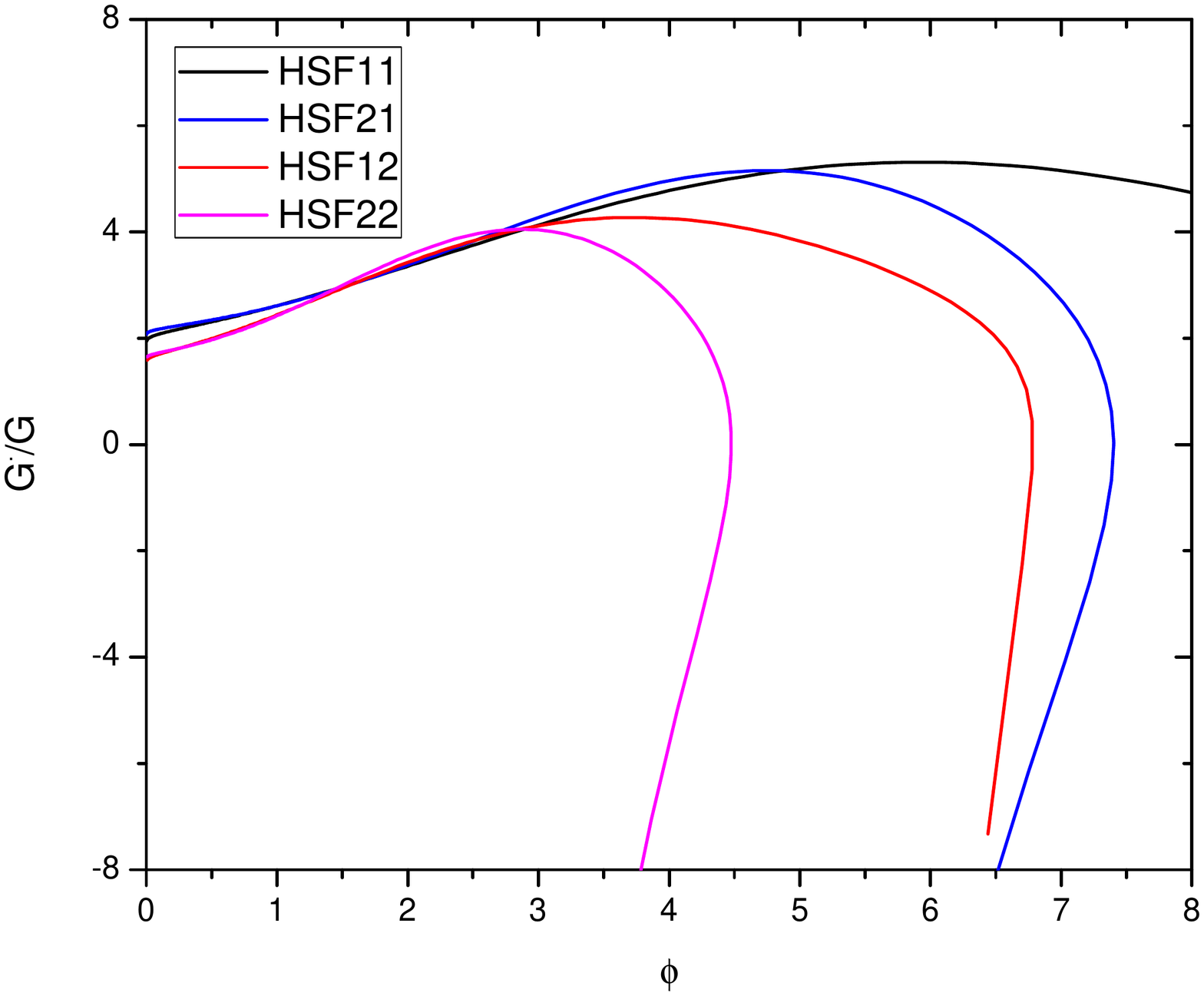}
\caption{ $\frac{\dot{G}}{G}$ as function of redshift $z$. The unified dark fluid parameter $\alpha$ is taken to be $0.00172$.}
\end{center}
\end{figure}
\subsection{Variation of Newtonian Gravitational Constant G}
In GBD theory, the Newtonian Gravitational constant $G(\phi)$ is expressed as
\begin{equation}
G(\phi)=\frac{4+2\omega(\phi)}{\phi(3+2\omega(\phi))},
\end{equation}
so that its time variation can be
\begin{equation}
\frac{\dot{G}}{G}=\frac{-2\dot {\omega}(\phi)}{(4+2\omega(\phi))(3+2\omega(\phi))}-\frac{\dot{\phi}}{\phi}.
\end{equation}

We have shown the time variation of $G$ as a function of redshift in Figure 11  for the four HSF models. The factor $\frac{\dot{G}}{G}$  increases from some large negative value in the past and exhibits a peak at some recent past ($z_{max} \sim 1$). Then it decreases to small positive value. $\frac{\dot{G}}{G}$ peaks at different redshifts for different models. The $z_{max}$ at which $\frac{\dot{G}}{G}$  peaks for a model depends both on the values of $z_{da}$ and $H^1$. For a given $H^1$, $z_{max}$ is higher for higher $z_{da}$. Similarly, for a given $z_{da}$, $z_{max}$ is higher for higher value of $H^1$.  The time variation of the Newtonian Gravitational constant is also shown as a function of the BD scalar field in Figure 12. For all the four HSF model, it is observed that, with an increase in $\phi$, the factor $\frac{\dot{G}}{G}$ increases initially and then decreases with $\phi$ after peaking at some value. The BD scalar field at which the peak of $\frac{\dot{G}}{G}$ will occur depends on the particular HSF models and respective model parameters. In order to have an idea about the order of magnitude of variation of $G$, we have calculated the value of $\frac{\dot{G}}{G}$ at the present epoch. The calculated values are given in Table II. For these calculation, we have taken the age of universe as $13.289\pm 0.289$ billion years\cite{Akarsu14}. For all the models $\frac{\dot{G}}{G}$ are in the order of  $10^{-10}yr^{-1}$. 

\section{Conclusion}

In the present paper, we have constructed some accelerating cosmological models in the frame work of generalised Brans-Dicke theory. We chose a Kantowski-Sachs universe that incorporates an anisotropy in the directional expansion rates. Kantowski-Sachs metric is a simple generalisation of FRW model. In the generalised Brans-Dicke theory, the Brans-Dicke parameter is considered to be evolving with the scalar field. Brans-Dicke theory is a modification of the GR where gravity is mediated through a dynamic scalar field. Observations have confirmed the late time cosmic speed up phenomena. In order to get accelerating models, we incorporate the concept of dark energy in the form of a unified dark fluid equation of state which takes care of both the dark matter and dark energy within a single equation. Constraints on the model parameters for the unified dark fluid from different observational data have been determined through different analysis. The range within which the model will work have been set up from the observational constraints. In order to get viable cosmological models within the frame work of GBD theory, we have assumed that the shear scalar is proportional to the scalar expansion which provides an anisotropic relation among the directional scale factors.

The universe is presently undergoing an accelerated expansion. It means that, at certain point of time, the universe may have undergone a transition from deceleration phase to an accelerated one. This behaviour requires a signature flipping of the deceleration parameter which should be positive at some early phase and negative at late phase of cosmic evolution. In order to simulate a time varying deceleration parameter with a behaviour of signature flipping, we consider a hybrid scale factor. The HSF has two factors.  While one factor dominates at an early phase of cosmic evolution the other dominates at late  phase. Therefore the HSF may be able to mimic a complete path history from an early deceleration to a late time acceleration. The parameters of the HSF need to be constrained to construct viable accelerating cosmological models. The transition redshift, at which the universe may have undergone a change from a deceleration to acceleration, may be considered as an important cosmological parameter. In view of this, we have explored the recently constrained transition redshift values to constrain the parameters of the HSF. Particulary, in the present work, we have used the values $z_{da}=0.8$  and $z_{da}=0.5$. Four different models namely, HSF11, HSF21, HSF12 and HSF22 are set up which provide accelerating models with the behaviour of an early deceleration to late time acceleration. The constructed models are within acceptable range of the observational $H(z)$ data and therefore consistent with observations. The deceleration parameter as predicted from these four constructed models lies in the range $-0.494 \geq q\geq -0.723$. These values are quite consistent with the recent estimates from observations. For these four accelerating models, we have investigated the evolution of the scalar field, Brans-Dicke parameter, Self interacting potential and the time variation of the Newtonian gravitational constant.

The BD scalar field, for all the constructed models, decreases from some large values at early phase to small values at the late epoch.  At a late phase of evolution, the BD scalar fields of all the HSF models behave alike. However, for the models HSF21 and HSF22  with $H^1=0.2$, $\phi$ first increases at an initial phase to a certain maximum and than decreases to the usual behaviour at some low redshift. The other two models with $H^1=0.3$, the evolutionary behaviour of the BD scalar field is bit smooth compared to the models with $H^1=0.2$. This behaviour is reflected in the evolutionary behaviour of other properties such as the BD parameter and the variation of the Newtonian gravitational constant. The BD parameter, for all the constructed models, increases with an increase in the BD scalar field. However at a late phase of evolution, the BD parameter practically becomes a constant quantity that may hint towards the convergence of the GBD theory to the usual Brans-Dicke theory. The self interacting potential for the constructed HSF models are attractive in nature and increases from some large negative value at an early epoch to vanishingly small values at late times of evolution. The four models behave alike in low redshift region where the scalar field assumes a smaller value. However, in the high redshift region with large scalar field, the four models splits to behave differently. The time variation of Newtonian constant is assessed through the behaviour of the quantity $\frac{\dot{G}}{G}$. It increases from some large negative value in the past and exhibits a peak at some recent past. Then it decreases to small positive value. $\frac{\dot{G}}{G}$ peaks at different redshifts for different models. The $z_{max}$ at which $\frac{\dot{G}}{G}$  peaks for a model depends both on the values of $z_{da}$ and $H^1$. The four HSF models constructed in the present work are consistent with the observational data. However, basing upon the behaviour of the BD scalar field, BD parameter and Self interacting potential, the two models HSF11 and HSF12 are more favourable as compared to the other models for cosmological studies. 

\section*{Acknowledgement}
SKT, DB and BM thank IUCAA, Pune (India) for providing support during an academic visit during which a part of this work is accomplished.

\end{document}